\documentclass[%
 reprint,
 superscriptaddress,
amsmath,amssymb,
prb,
]{revtex4-1}
\usepackage{graphicx}
\usepackage{dcolumn}
\usepackage{bm}

\begin{document}

\title{Density functional theory calculations and thermodynamic analysis of the forsterite $Mg_{2}SiO_{4}$(010) surface}

\author{Ming Geng}
\email{gengming@mail.iggcas.ac.cn}
\affiliation{Key Laboratory of Earth and Planetary Physics, Institute of Geology and Geophysics, Chinese Academy of Sciences, Beijing 100029, China}
\affiliation{Institutions of Earth Science, Chinese Academy of Sciences, China}
\author{Hannes J\'{o}nsson }
\affiliation{Faculty of Physical Sciences, University of Iceland, 107 Rekjav\'{i}k, Iceland}
\affiliation{Dept. of Energy Conversion and Storage, Technical University of Denmark, DK-2800 Kgs. Lyngby, Denmark}


\begin{abstract}
The stability of possible termination structures for the (010) surface of forsterite, $ Mg_2SiO_4 $, is studied using a density functional theory (DFT) based thermodynamic approach. The DFT calculations are used to estimate the surface Gibbs free energy of various surface structures and compare their stability  as a functions of the chemical environment. Among 9 possible terminations, the SiO-II, M2, O-II terminations are found to be most stable as conditions range from Mg-poor to Mg-rich.  This relative stability order remains the same at elevated temperature. The surface phase diagram obtained provides ground for further theoretical studies of chemical processes on forsterite surfaces in terrestrial planets.

\end{abstract}

\maketitle

\section{\label{sec:level1}Introduction}

Silicate minerals are the building blocks of terrestrial planets. Within a wide variety of silicate minerals, olivine is the predominant mineral in both Earth’s upper mantle and interstellar media in space. 
Consequently, olivine plays a fundamental role in defining the properties and influencing the physiochemical processes of the interior of terrestrial planets. Knowledge of physical and chemical propertied of olivine is of great geophysical and astrophysical interest because they play an important role in many important processes.

\begin{figure*}[h]
	\centering
	\includegraphics[width=0.9\linewidth]{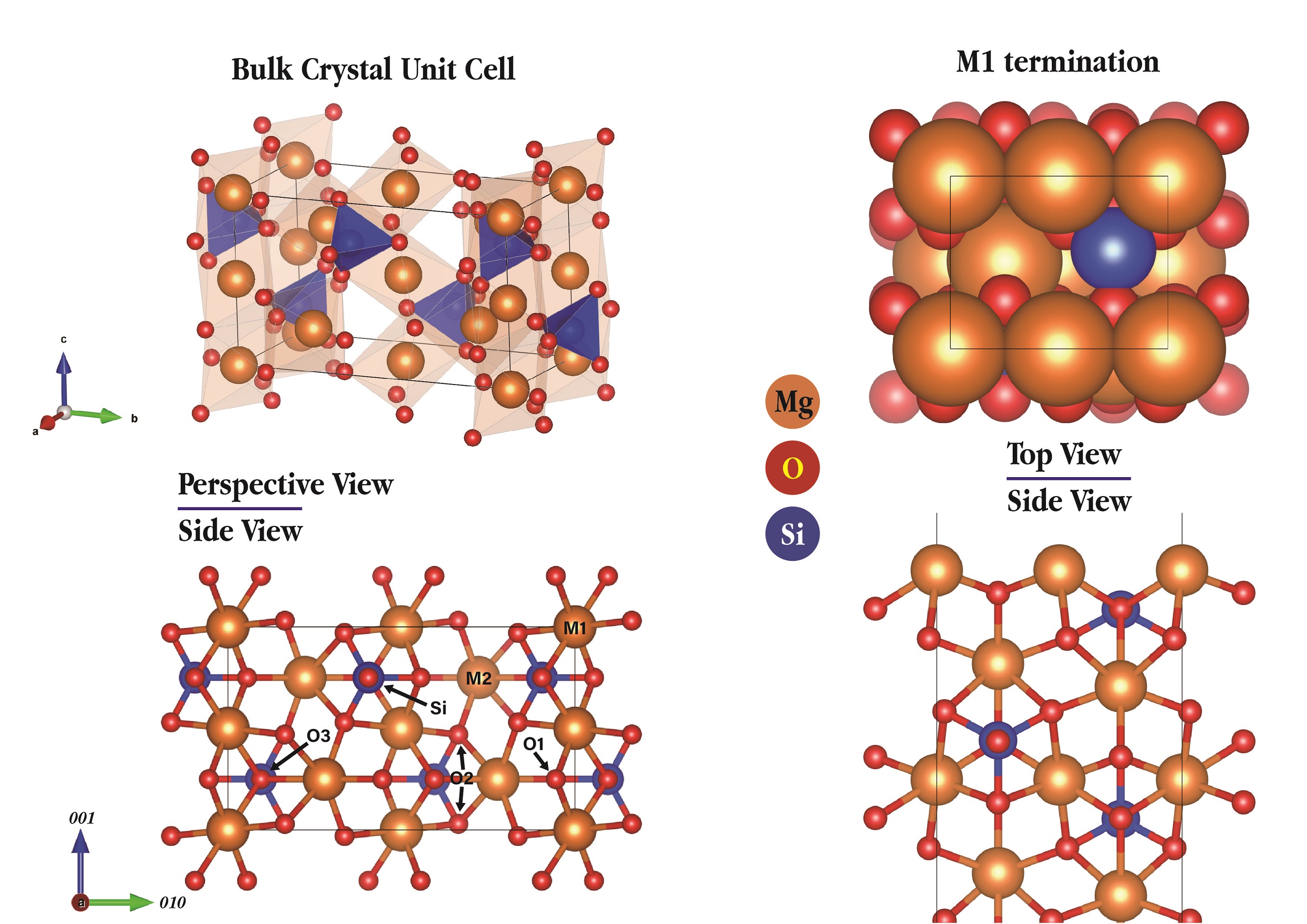}
	\caption{Structure of the forsterite unit cell  and M1 termination}
	\label{fig:M1}
\end{figure*}

Olivines are silicate solid solutions with the general formula $(Mg_xFe_{2-x}) SiO_4$.  Forsterite is at the Mg rich end of the range corresponding to $x=2$ and fayalite corresponds to the iron rich end with $x=0$. Both on Earth and in space, the most common olivines are richer in magnesium than in iron. Olivine crystals have orthorhombic structure with space group \textsl{Pbmn} and are characterized by $[SiO_4]^{4-}$ tetrahedra linked by the divalent metal cations as indicated in Fig.\ref{fig:M1}. 

The exposed surface structure or termination of forsterite $Mg_2SiO_4$ is determines its ability to adsorb volatile chemicals, an important aspect in many natural processes. For example, the chemisorption of water molecules onto forsterite dust grain has been considered the possible source of water on terrestrial planets\cite{RN2702,RN2699,RN2703,RN2707}. Despite its low porosity and permeability, the formations of the ultramafic rock peridotite provides attractive sites for permanent carbon dioxide sequestration because of the high olivine content and thus significant potential for carbon dioxide mineralization \cite{RN298242,RN298394}. During the subduction of a tectonic plate, carbonated peridotite (Olivine + $ CO_2 $) and serpentinization (Olivine + $ H_2O $) can transport carbon and water into the deep interior of the Earth. This process is an important pathway of the volatile element cycle on Earth and affects many important features of our planet. Therefore, a fundamental understanding of the surface structure of $Mg_2SiO_4$ under various reaction conditions is a prerequisite for atomic scale understanding of various processes on Earth.

Several computational studies using empirical force fields  \cite{RN2686,RN2700}  have been conducted of the main low-index surface of forsterite. Recently, {\it ab initio} calculations \cite{RN298173,RN2646,RN2684, RN2685} with various basis sets have also been performed. Both the force field and {\it ab initio} calculations have shown that the (010) surface of forsterite is  the most stable one among the various low-index surfaces. Several studies have also been conducted of the absorption of volatile atoms and molecules, such as H atom \cite{RN2684} and water molecule \cite{RN2656}. All these studies have been based on the non-polar surfaces.Polar surface terminations may, however, also be important. An important issue is surface termination, i.e. which atoms are exposed at the (010) surface.Previous studies, have not been based on the same termination and apparently it has been selected rather arbitrarily. 

In recent experimental work based on high-resolution X-ray reflectivity (HRXR)\cite{RN2636} measurements, various surface terminations and different levels of hydrations of the forsterite(010) surface were observed depending on the sample preparation such as the method used for polishing the surface. The termination is, therefore, an important consideration when determining the relative stability of forsterite surfaces and in studies of molecular adsorption.

In this article, we present simulations of $Mg_2SiO_4$(010) surfaces with all possible terminations and analyze their structure and surface stability as a function of oxygen partial pressure at high and low temperature using density functional theory (DFT) calculations and thermodynamics.

Section II describes the computational methodology and the surface structures. In section III the results obtained from the calculations are presented along with some  discussion. Conclusions are summarized in Section IV.


\section{\label{sec:level1} METHODOLOGY}

\subsection{\label{sec:level2}Surface structures}

There are two different metal ion sites, three oxygen sites and one silicon site in forsterite, $Mg_2SiO_4$. In the [010] surface orientation, the atomic layers are stacked $  –2Mg(M1)-O(O1)-SiO(O3)-2O(O2)-Mg(M2)-Mg(M2)-2O(O2)-SiO(O3)-O(O1)-2Mg(M1)- $. Crystal planes with [010] orientation can have 9 different surface terminations. In the present study, slabs are constructed with these nine structurally different (010) surface terminations. The slabs are labeled after the cut positions of the atomic stack as follows: 
M1 (metal site 1), $ SiO$ (O3 and Si site), $ O $ (O1 site), $ O2 $ (O2 site)  and M2 (metal site 2). 
Because  M1 (metal site 1) is on a symmetry line, it can have only one surface structure. The other four cut sites ($ SiO $, $ O $, $ O2 $ and M2) can form two different surface structures in opposite directions. We use a "-II" mark to distinguish between different slabs in the same position.  Each slab has the same termination on both sides. Of these nine slabs, only the M2 terminated one is stoichiometric.  


\subsection{\label{sec:level2}Cleavage Energy}
When a forsterite crystal is cleaved, two complementary surfaces are created. There are five complementary pairs of surfaces for the forsterite [010] orientation. The M2-terminated surface is complementary to itself, and the created slab is stoichiometric. In the present study, asymmetric slabs terminated with different surfaces on two sides are used. Since the combination of two slabs with complementary terminations is a two-time-thick unit bulk crystal, the cleavage energy per are of a unit cell is calculated as follows: 

\begin{equation}
\label{cleavageEnergy}
E_{cleave}(i, j) = - \frac{1}{2}[E^i_{slab}+E^j_{slab}-2E^{bulk}_{Mg_2SiO_4}]
\end{equation}

where $E^{bulk}_{Mg_2SiO_4}$ is the total energy for a unit cell of $Mg_2SiO_4$, and $E^i_{slab}$ and $E^j_{slab}$  are the values of the total energy for complementary surface slabs. The cleavage energy density per unit area is equal to 

\begin{equation}
\label{cleavageEnergyDensity}
\epsilon_{cleave}(i,j)=\dfrac{E_{cleave}(i,j)}{A}
\end{equation}
where $A$ is the surface area of a unit cell.

\subsection{\label{sec:level2}Computational method}

The DFT calculations make use of the Perdew-Burke-Ernzerhof (PBE) approximation to the exchange-correlation functional \cite{RN298236} with a 500 eV kinetic energy cutoff in the plane wave basis set.  The projector-augmented wave (PAW) method\cite{RN298234} is used to describe the effect of the inner electrons. Monkhorst-Pack k-point mesh schemes were used in the optimization for bulk phases (8$\times$8$\times$8) and surface terminations (8$\times$8$\times$1). To minimize the effect of the periodic boundary condition, a 30 \AA-thick vacuum space is added on top of the surface slab models. The ground-state atomic geometries of the bulk and surface are obtained by minimizing the forces on each atom to below 0.01 eV/\AA. The Vienna Ab-initio Simulation Package (VASP) code \cite{RN298230,RN298231,RN298232,RN298233} is used to perform the DFT calculations.

\subsection{\label{sec:level2}Thermodynamic stability}

The results of the DFT calculations are used to estimate the thermodynamic stability of the various forsterite surfaces in equilibrium with $ O_2 $ and the various oxides. By evaluating the Gibbs free energy, the exchange of atoms between the bulk crystal, its surface and the gas phase is taken into account in the analysis of surface stability. The surface Gibbs free energy is a measure of the excess energy of a semi-infinite crystal in contact with matter reservoirs with respect to the bulk crystal. The most stable surface is the one with the smallest surface Gibbs free energy.

The surface Gibbs free energy is a function of the chemical potentials of the various atomic species. For example, the surface Gibbs free energy of a forsterite slab is:
 \begin{equation}  \label{GDefine}\Omega _i = \dfrac 1{2}(G^i _{slab} - N_{Mg} \mu _{Mg} - N_{Si} \mu_{Si} -N_O \mu_{O}) \end{equation}
 where $N_{Mg}$ ,$N_{Si}$, $N_O$ denote the numbers of Mg, Si and O atoms in the slab, $G^i _{slab}$ is the Gibbs free energy of the symmetric slab with identical surface terminations on both top and bottom sides, and $\mu_{Mg}$, $\mu _{Si}$ and $\mu _O$ are the chemical potentials for the Mg, Si and O atomic species in the forsterite crystal. The surface Gibbs free energy per unit area is
 
\begin{equation}   \label{GperS} \omega = \dfrac{\Omega}{A} \end{equation}
 
Because the surface of each slab is in equilibrium with bulk forsterite, the chemical potential of $Mg_2SiO_4$ is equal to the bulk crystal Gibbs free energy, which results in the following expression:
 
\begin{equation} \label{miusum} \mu _{Mg_2SiO_4}=g^{bulk}_{Mg_2SiO_4}\end{equation}

where $\mu _{Mg_2SiO_4}$ is the chemical potential of forsterite, which is equal to the sum of the chemical potentials of all atom types in the $Mg_2SiO_4$ crystal:
 
 \begin{equation} \label{SurfaceGibbs}
 	\mu _{Mg_2SiO_4} = 2\mu_{Mg} + \mu_{Si} +4\mu_{O}
 \end{equation}

The equation for the surface Gibbs free energy as a function of the variations of the Mg chemical potential and O chemical potential can be simplified as

\begin{equation}\label{ExcessAtom}
\Omega^i = \dfrac{1}{2} (G^i_{slab}-N^i_{Si}g^{slab}_{Mg_2SiO_4}) - \Gamma^i_{Si,Mg} \mu_{Mg} - \Gamma^i_{Si,O}\mu_{O}
\end{equation}

where the parameters $ \Gamma^i_{Si,Mg}$ and $\Gamma^i_{Si,O}$ are the excesses in the $i$ terminated surface of Mg and O atoms with respect to the number of Si ions in the slab:

\begin{equation}
	 \Gamma^i_{Si,Mg} = \dfrac{1}{2}(N^i_{Mg}-N^i_{Si}\dfrac{N^{bulk}_{Mg}}{N^{bulk}_{Si}})
\end{equation}

In order for the Mg and Si atoms and $O_2$ molecule not to have a thermodynamical driving force to leave the bulk  $Mg_2SiO_4$ crystal and precipitating on the surface, their chemical potential must be less than the Gibbs free energy of the corresponding bulk \cite{RN2693}:
\begin{equation} \label{BoundaryMg}\mu_{Mg} \le g^{bulk}_{Mg}\end{equation}
\begin{equation}\label{BoundarySi}\mu_{Si} \le g^{bulk}_{Si}\end{equation}

Similarly, the precipitation of $MgO$ and $SiO_2$ will not be thermodynamically favored if
\begin{equation}\label{BoundaryMgO}\mu_{Mg}+\mu_{O} \le g^{bulk}_{MgO} \end{equation}
\begin{equation}\label{BoundarySiO2}\mu_{Si}+2\mu_{O} \le g^{bulk}_{SiO_2} \end{equation}

The vibrational contribution to the surface Gibbs free energy is often neglected \cite{RN2747}, but it can make a significant contribution when surfaces with similar stabilities are compared. 
We explicitly introduce the vibrational contribution to the Gibbs free energy.  The Gibbs free energy $ g $ can be approximated as the total energy and $F^{vib}$:

\begin{equation}
\label{Gibbs}
g=E+F^{vib}-TS^{conf}+pV \approx E+F^{vib} 
\end{equation}
Here, the terms $F^{vib} $ ,$T$, $ p $, $ V $ and $S^{conf}$ correspond to the vibrational free energy,  temperature, pressure, volume, and configurational entropy, respectively.  In addition,  the crystal volume per molecule is approximately 73.174  \AA $ ^3 $, so the largest $pV$ term is only several meV for 100 atm pressure, which is negligible compared with typical error bars in the DFT calculations.
$TS^{conf}$ is also approximated to be zero because of its negligible energetic contributions, as has been done in previous studies. Thus, $g$ in Equation \ref{Gibbs} is simplified to have contributions from only $ E $ and $ F^{vib} $. The $F^{vib}$ term can be expressed using the phonon density of states (DOS) as follows:

\begin{equation}\label{Phonon}
F^{vib}= \dfrac{1}{2}\sum\limits_{qj} h \omega_{qj} + k_{B}T \sum_{qj} \ln[1- \exp(-h \omega_{qj} / k_{B}T)]
\end{equation}
where $q$ is the wave vector, $j$ is the band index, and $\omega_{qj}$ is the phonon frequency of the phonon mode labeled by a set \{q,j\}. The phonon calculations were conducted based on the density functional perturbation theory (DFPT) implemented in the VASP/Phonopy software \cite{RN2771}.

By introducing the deviation in the Mg and Si chemical potentials
\begin{equation} \label{dmiuMg}
\Delta \mu_{Mg} = \mu_{Mg} - g^{bulk}_{Mg} 
\end{equation}
and
\begin{equation}\label{dmiuSi}
	\Delta \mu_{Si} = \mu_{Si} - g^{bulk}_{Si}
\end{equation}
Then,  Equation(\ref{miusum}) can be rewritten as
\begin{equation}\label{dmiusum}
	2\Delta \mu_{Mg} + \Delta \mu_{Si}+ 4\Delta \mu_{O} = \Delta g_f(Mg_2SiO_4)
\end{equation}
where  $\Delta g_f(Mg_2SiO_4)$ is the formation Gibbs free energy of $Mg_2SiO_4$ from Mg, Si and $O_2$ in their standard states.
\begin{equation}\label{formationG}
	\Delta g_f(Mg_2SiO_4) = g^{bulk}_{Mg_2SiO_4} - 2g^{bulk}_{Mg} - g^{bulk}_{Si} - 2E_{O_2}
\end{equation}
The above boundary conditions can be transformed into:
\begin{equation} \label{dBoundaryMg}	\Delta \mu_{Mg} \le 0  \end{equation}
\begin{equation}\label{dBoundarySi}  2\Delta \mu_{Mg} + 4 \Delta \mu_{O} \ge \Delta g_f(Mg_2SiO_4) \end{equation}
\begin{equation}\label{dBoundaryMgO} \Delta \mu_{Mg} +\Delta \mu_{O} \le \Delta g_f(MgO) \end{equation}
\begin{equation}\label{dBoundary4SiO2}\Delta \mu_{Mg} + \Delta \mu_{O} \ge \frac{1}{2}( \Delta g_f(Mg_2SiO_4)-\Delta g_f(SiO_2))\end{equation}

where $\Delta g_f(MgO)$ and $\Delta g_f(SiO_2)$  are the Gibbs free energies of formation for $MgO$ and $SiO_2$. The oxygen atoms in $Mg_2SiO_4$ are in equilibrium with oxygen gas in the atmosphere above the crystal surface:
\begin{equation}\label{miuO}	\mu_{O} =\frac{1}{2} \mu^{gas}_{O_2}  \end{equation}

The chemical potential of oxygen gas  depends on the gas temperature and partial pressure. This dependence can be used to express the Gibbs free energy of the $ Mg_2SiO_4 $ surface in terms of the temperature and oxygen gas partial pressure. The oxygen gas as is approximated as an ideal gas, so the dependence of the chemical potential on pressure can be written as

\begin{equation}
\begin{aligned}\label{miuOgas}
\mu^{gas}_{O_2}(T,P)&=\mu^{gas\ominus}_{O_2}(T)+kT\ln f_{O_2}(T,P)\\
&=\mu^{gas\ominus}_{O_2}(T)+kT\ln(\frac{P}{P^0})
\end{aligned}
\end{equation}
where $k$ is the Boltzmann constant,
 $\mu^{gas\ominus}_{O_2}(T)$ is the standard chemical potential of oxygen gas at temperature T, evaluated from the NIST-JANAF data\cite{RN99999}. The variation of the oxygen atom chemical potential can be written as

\begin{equation}
\begin{aligned}\label{dmiuO}
\Delta \mu_{O}(T,P)&= \mu_{O}(T,P)-\frac{1}{2}E_{O_2}\\&=\frac{1}{2} [\Delta G^{gas}_{O_2}(T,P^0)+kT\ln(\frac{P}{P^0})]+\delta \mu ^0 _O
\end{aligned}
\end{equation}

where
 $\Delta G^{gas}_{O_2}(T,P^0)$ is the change in oxygen gas Gibbs free energy at standard pressure $ P^0 $ and temperature T with respect to its Gibbs energy at 298.15 K. We present the calculated values for $\Delta G^{gas}_{O_2}(T,P^0)$  in Table \ref{tab:OxygenGibbs}. $\delta \mu ^0 _O$  is a correction, to match the DFT calculation of O$_2$ (which is known to have a relatively large error) to experimental data.  This correction was estimated from the computation of metal oxides and metals similarly to the approach in Reuter and Scheffler\cite{RN2747}. In our calculation, $\delta \mu ^0 _O$ is 0.1997 eV and 0.4208 eV for $ SiO_2 $ and $ MgO $, respectively. We use an average value of 0.3102 eV for the calculation hereafter. 
 
 Finally, we write Equation \ref{Gibbs} as 
 
 \begin{equation}\label{Gdmiu}
 	\Omega^i = \phi ^i - \Gamma ^i _{Si, Mg}\Delta \mu_{Mg} - \Gamma ^i _{Si,O}\Delta \mu_O
 \end{equation}
 
 and
 $\phi ^i$ as:
  
   \begin{equation}\label{phi}
\begin{split}
\phi ^i&=\frac{1}{2}(G^i_{slab}-N^i_{Si}g^{bulk}_{Mg_2SiO_4})-\Gamma^i_{Si,Mg}g^{bulk}_{Mg}-\frac{1}{2}\Gamma^i_{Si,O}E_{O_2} \\
&\approx\frac{1}{2}(E^i_{slab}+F^{vib}_{slab}- N^i_{Si}g^{bulk}_{Mg_2SiO_4}) \\
  &   -\Gamma^i_{Si,Mg}g^{bulk}_{Mg}-\frac{1}{2}\Gamma^i_{Si,O}E_{O_2}
\end{split}
  \end{equation}
  
The surface Gibbs free energy for all surfaces described in Section II A was calculated. The region where the $Mg_2SiO_4$ surface is stable with respect to the precipitation of Mg, Si or their oxides is determined using the inequalities.

\section{\label{sec:level1}Results and Discussion}

\subsection{Atomic and electronic structure}

The optimized lattice constant of the bulk crystal obtained from the DFT calculation is close to what has been
reported from previous calculation \cite{RN2646,RN298257} and experimental measurements \cite{RN298259}, as summarized in Table \ref{tab:LatticeConstant}. 

\begin{figure*}[h]
	\centering
	\includegraphics[width=0.9\linewidth]{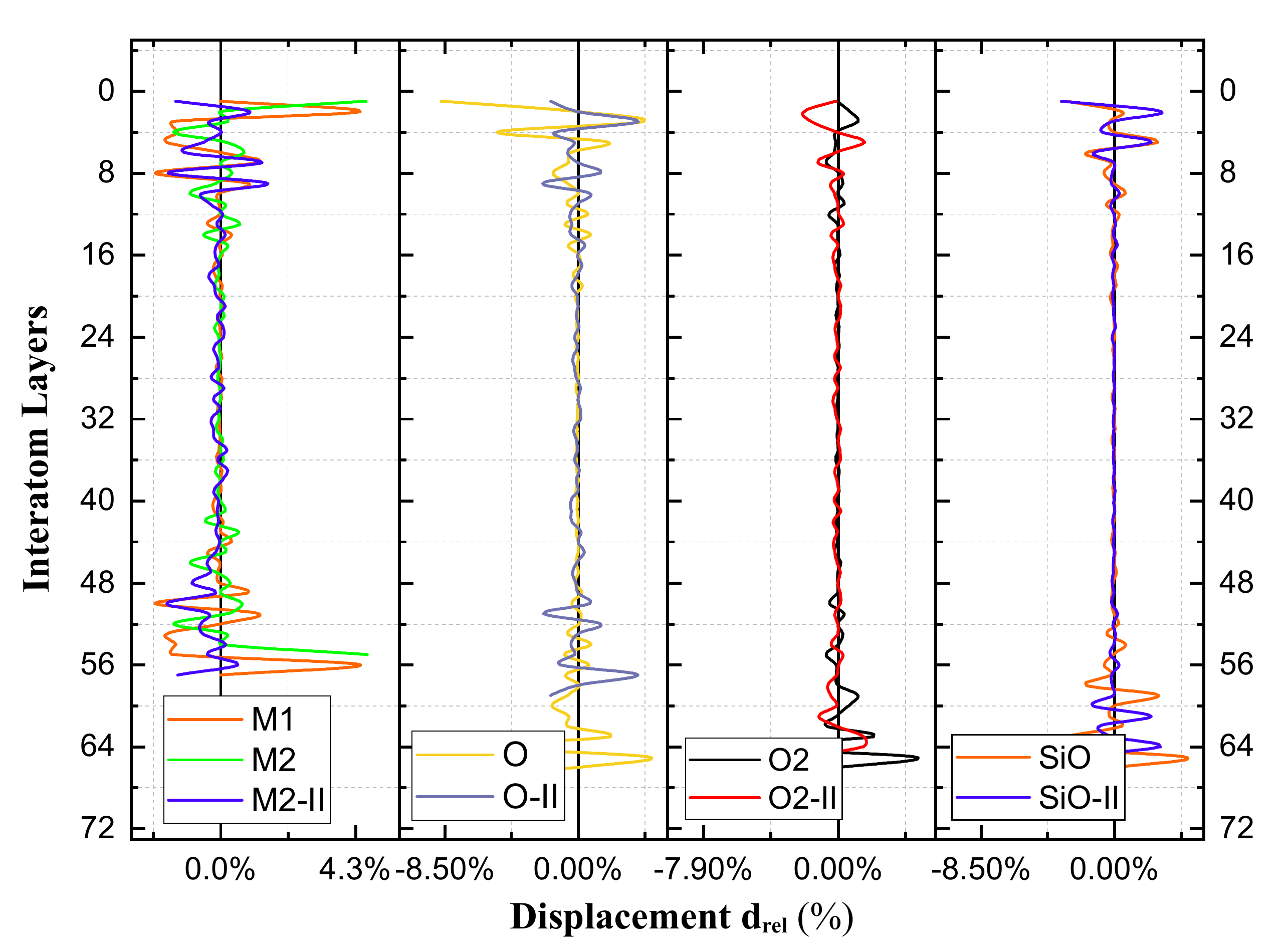}
	\caption{Displacement oscillation pattern of the relaxed atom layers}
	\label{fig:Displacement}
\end{figure*}

After the relaxation of the slab surface, the topmost atoms are displaced from the ideal crystal positions towards the interior of the slab.  These displacements are listed in Table \ref{tab:Displacement}, and are evaluated as

\begin{equation}\label{dispalcement}
d_{rel}=\dfrac {\Delta Z^{origin}_{i,j}-\Delta Z^{relaxed}_{i,j}}{\Delta Z_0} .
\end{equation}
Here, $\Delta Z^{origin}_{i,j}$ is the distance between atoms $i$ and $j$ in the unrelaxed slab, $\Delta Z^{relaxed}_{i,j}$ is the distance between atoms $i$ and $j$ in the relaxed slab, and $\Delta Z_0$ is the length of the unit cell of the forsterite crystal. For $d_{rel}$, a positive sign implies a contraction between the surface layers, whereas a negative sign implies an expansion. 
The slab models are thicker than twice the bulk crystal cell (more than 40 layers), and the displacement of atoms from the crystal position is confined only to layers close to the surface. 
Figure \ref{fig:Displacement} shows that the displacement becomes negligible beyond 10 layers. A unit cell of bulk crystal forsterite has 19 layers in the [010] direction.  In M1, O2 and O2-II terminations, the topmost two layers have the same types of atoms. These atoms remain in the same layer after relaxation. In almost all terminations, the displacement between O and Mg atoms is larger than the displacement between O and Si or O layers. For example, in the M1 and M2 terminations, the Mg atoms in the topmost layers are displaced towards the bulk as indicated by the large positive $d_{rel}$. The O atom in the topmost layer of the O and O-II termination is less constrained. An interlayer expansion appears in the first two layers of these two terminations. Si and O are also split into two layers at the topmost SiO and SiO-II terminations.

The calculated Bader charges \cite{RN298410} for the atoms in the outermost few layers of all nine terminations are listed in Table \ref{tab:Bader}. For reference, the charges of Mg, Si, and O atoms in  $Mg_2SiO_4$ crystal are also listed in Table~\ref{tab:Bader}. 
The M2 , O-II and SiO-II terminations have the smallest charge difference among these nine terminations. The variations are all not larger than 0.05 e in the first 10 layers, as shown in the Table. \ref{tab:Bader}. The other six terminations have similar charge variations of approximately 1.50 e in the first 10 layers.

\subsection{Cleavage energy}
The energy required to create complementary surface pairs for forsterite (010) were calculated using 
equations (\ref{cleavageEnergy}) and (\ref{cleavageEnergyDensity}), and the results are collected in Table \ref{tab:CleavageEnergy}. The largest cleavage energy was obtained for a pair of O2- and SiO-terminated (010) surfaces. The lowest energy is the creation of an M2-M2 terminated (010)  surface pair. This slab is also the only stoichiometric slab. Therefore, one should expect the formation of these surfaces, when an $Mg_2SiO_4$ crystal is cleaved perpendicular to the [010] direction. Thus, the M2 termination is the one previous forsterite surface-related studies have focused on \cite{RN2637,RN2645,RN2684}.  Evolution of these surfaces from one termination to another depends on the possibility of ion exchange between crystal and surface, surface and environment, and mobility of atoms on the surface.

\subsection{Surface stability of various terminations}
\begin{figure*}[h]
	\centering
	\includegraphics[width=0.8\linewidth]{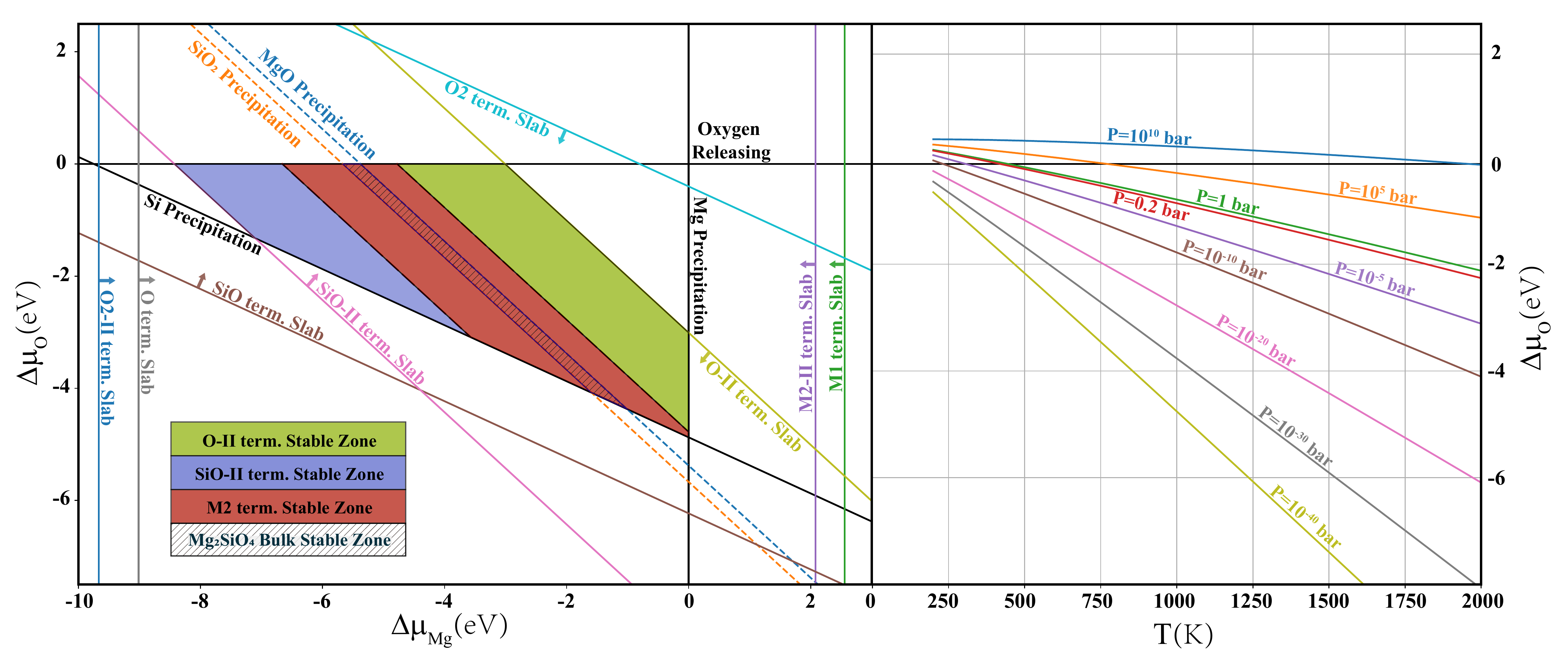}
	\caption{Phase diagram: the regions of stability of the $ Mg_2SiO_4 $ (010) surfaces with different terminations as functions of the chemical potential variations for Mg and O atoms}
	\label{fig:Figure}
\end{figure*}

The cleavage energy is the energy required to split a crystal into two parts with complementary terminations. Therefore, the cleavage energy itself does not give directly the stability of the different surfaces formed.  The Gibbs surface free energy is a measure of the excess energy of a semi-infinite crystal in contact with matter reservoirs, and can be used to analyze the stability of  the various surface terminations. The DFT calculations have been used to determine all parameters needed to calculated the surface free energy of a variety of surfaces and formation energies for $ Mg_2SiO_4 $, $ MgO $ and $ SiO_2 $ crystals. The excess of oxygen and Mg atoms with respect to Si atoms in the simulated slabs and values are calculated and listed in Table \ref{tab:Excess_atoms}. A spontaneous surface formation line of every surface with a specific termination, presented on phase diagrams (shown on the left part of Fig. \ref{fig:Figure} with a label denoting the termination),  can be determined by solving the equation

\begin{equation}\label{formation Line}
	\Omega^i(\Delta \mu_{Mg}, \Delta \mu_{O}) = 0
\end{equation}

The direction indicated by the arrow on each spontaneous formation line, shows where the Gibbs free energy becomes positive, suggesting that these terminations are stable and may be exposed on $Mg_2SiO_4$ particles under the given conditions. The most stable surface for any particular value of the Mg and O chemical potentials is the surface with the smallest, positive surface Gibbs free energy. The boundaries among the regions of stability for different terminations are marked by red dashed lines. The boundary between stability regions for surfaces with terminations $i$ and $j$ is determined by the solution of the equation

\begin{equation}\label{Boundary Condition}
\Omega^i(\Delta \mu_{Mg}, \Delta \mu_{O}) =\Omega^j(\Delta \mu_{Mg}, \Delta \mu_{O})
\end{equation}

 In the present study, only three surface terminations satisfy the two condition at the same time. Thus, we only plot two boundaries of these three surfaces. On the left part of Fig.\ref{fig:Figure}, each color block represents the region of stability for a different surface termination.\par
 
Pure $Mg_2SiO_4$ exists when the conditions of Equations (\ref{dBoundaryMg}) -  (\ref{dBoundary4SiO2}) are all satisfied. All of these conditions are shown in Fig. \ref{fig:Figure} as solid lines. The formation energies of $Mg_2Si_4$,  magnesium and silicon oxides determine the position of respective precipitation lines. Precipitation of silicon occurs below the Si precipitation line, magnesium metal precipitates on the right from the magnesium precipitation line. $ MgO $ crystal will grow on the right and above the $ MgO $ precipitation line, and $ SiO_2 $ will grow on the left from the $ SiO_2 $ precipitation line. A pure $ Mg_2SiO_4 $ can only be obtained in the narrow striped region between the $ MgO $ precipitation line on the left and the $ SiO_2 $ precipitation line on the right. At the bottom of the diagram, the strip is limited by the $ Si $ precipitation line.  
  
The calculated stability diagram is presented in Fig. \ref{fig:Figure}. It shows regions of oxygen and magnesium chemical potentials where the surface free energy [Equation(\ref{Gdmiu})] are calculated for the various surface terminations are minimal. Every color block is a stability zone of the corresponding surface slab.

At each point on the diagram, the $ Mg_2SiO_4 $ surface is in equilibrium with oxygen gas. The equilibrium is characterized by the oxygen chemical potential. Since a single value of the chemical potential can correspond to a wide range in temperature and pressure,  the dependency of the oxygen chemical potentials on temperature is shown for a number of values for the gas pressure on the right side of Fig.\ref{fig:Figure}. These functions were calculated from experimental data, taken from the thermodynamical tables \cite{RN99999} following the approach described earlier by Equation (\ref{dmiuO}). The design used for constructing the diagrams in Fig. \ref{fig:Figure} make it possible to determine the conditions for the oxygen environment that correlates with the points on the phase diagrams on the left side of the figures.

\begin{figure*}
	\centering
	\includegraphics[width=0.8\linewidth]{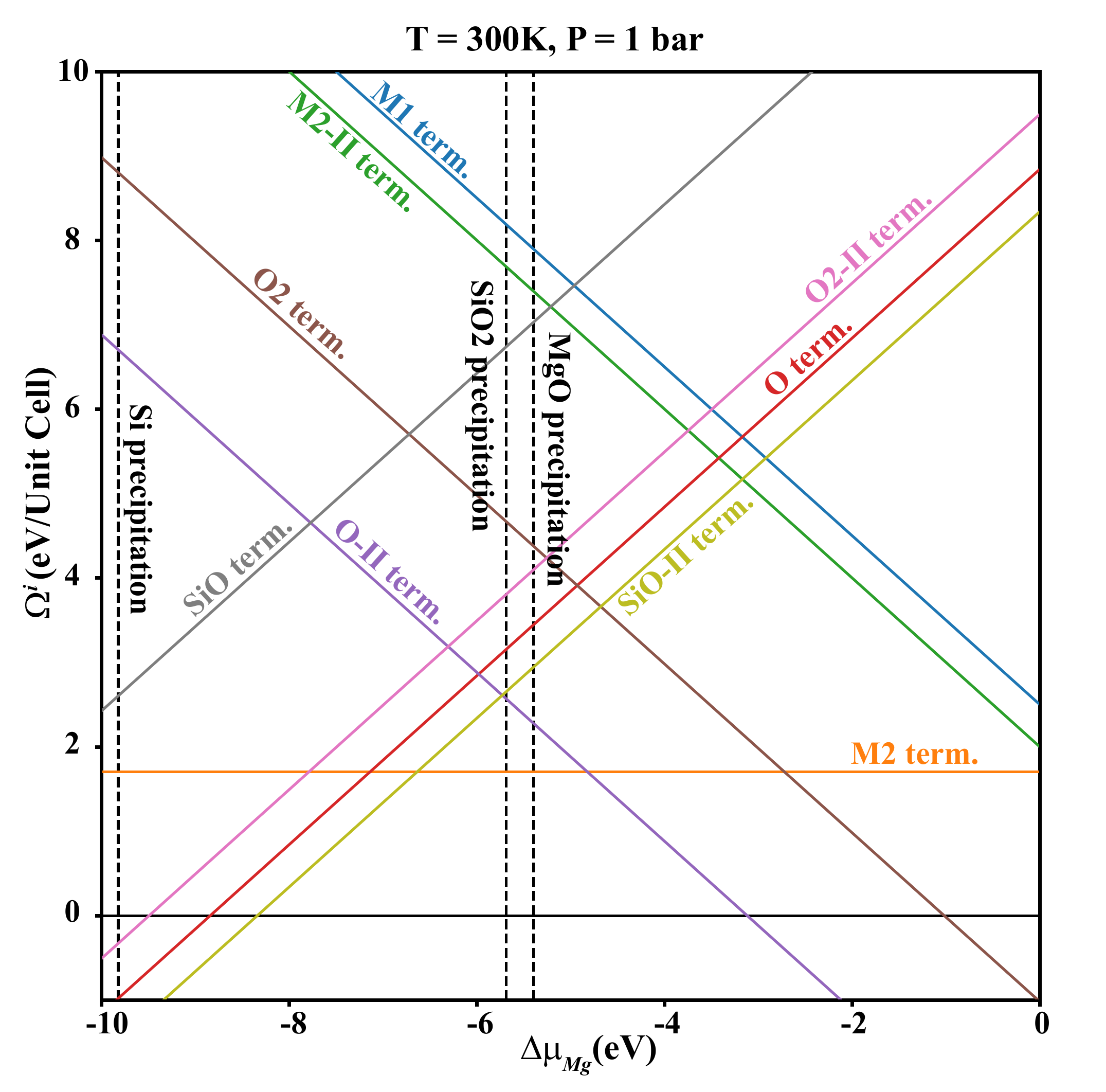}
	\caption{Surface Gibbs free energy as a function of $\Delta \mu _{Mg}$ at T=300 K and $p_{O_2} = 1 bar $}
	\label{fig:Surface free energy 300}
\end{figure*}

\begin{figure*}
	\centering
	\includegraphics[width=0.8\linewidth]{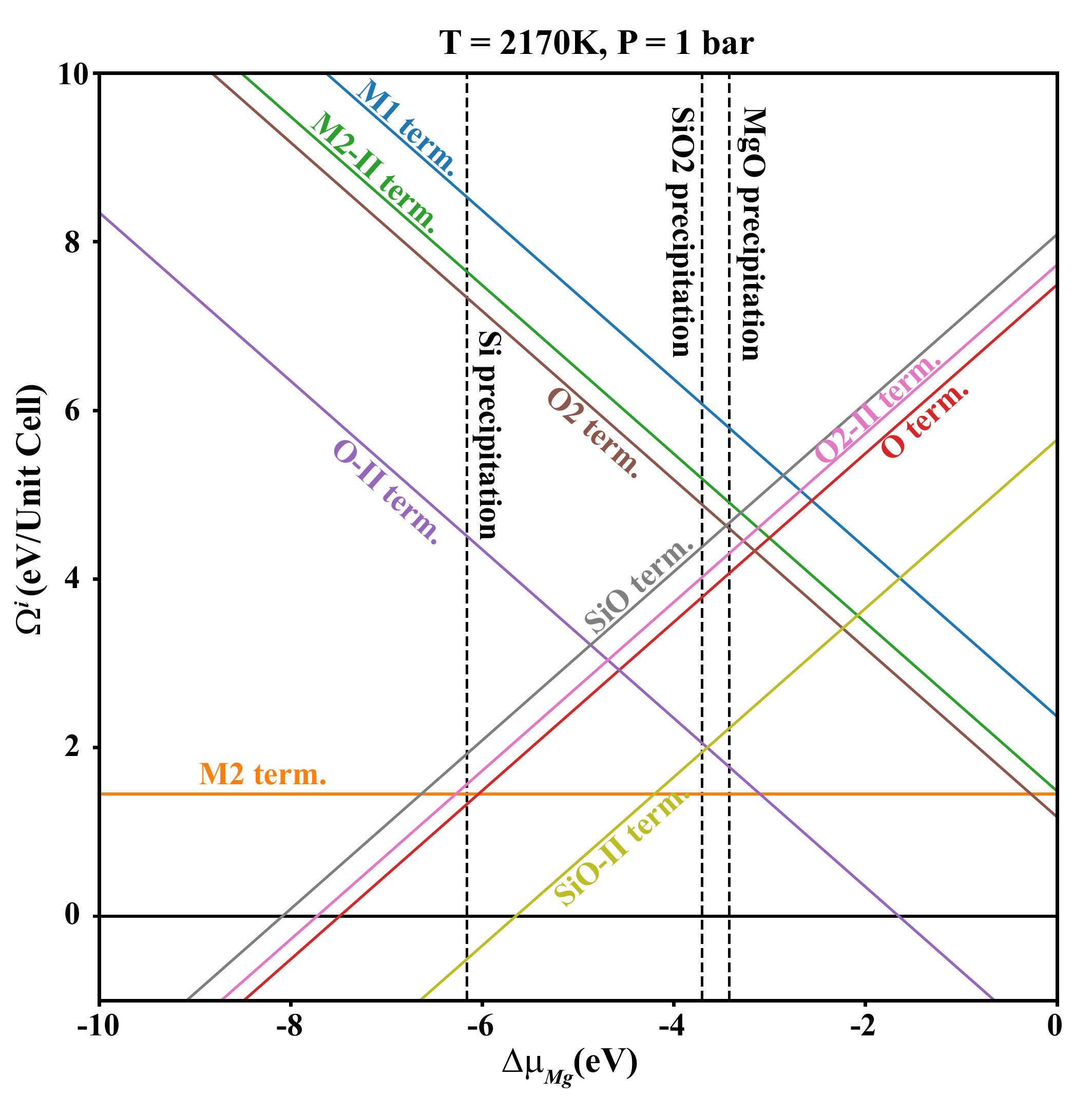}
	\caption{Surface Gibbs free energy as a function of $\Delta \mu _{Mg}$ at T=2170 K and $p_{O_2} = 1 bar $}
	\label{fig:Surface free energy 2170}
\end{figure*}

To illustrate the calculated results for the surface Gibbs free energy of the various $Mg_2SiO_4$ surfaces, we show graphs for several specific conditions.  The surface Gibbs free energy defined in Eq. \ref{Gdmiu} for an oxygen gas pressure equal to 1 bar is shown in Fig. \ref{fig:Surface free energy 300} and Fig. \ref{fig:Surface free energy 2170} representing ambient conditions and near-melting conditions\cite{RN298311}, respectively. For smaller $\Delta \mu_{Mg}$ (corresponding to Mg-poor conditions), the most stable surface is SiO-II terminated. When the environment is Mg-rich, the M2,+ and O-II terminated surfaces become more stable.  As the temperature is increased, the ordering of the stability of these surface terminations is unchanged.


\section{\label{sec:level1}Conclusion}
We used DFT calculations and thermodynamic methods to analyze nine possible surface terminations of the forsterite(010) surface. The relative stability of the various surfaces has been estimated as a function of magnesium and oxygen chemical potentials.  While most previous studies have focused on the M2 terminated surface, we find that the SiO-II and O-II surface can also be stable under some conditions. Especial the O-II termination shows a possible exposure of M1 site on the surface. Previous study\cite{RN2847} shows the Fe has strong preference to occupy the M1 site in olivine. And Fe in M1 is not been focus on in the olivine surface reaction  researches.
Because the cleavage energy for creating a pair of M2 terminated surfaces is smaller than the cleavage energy for creating other pairs of forsterite(010) surface, it will is most likely to observe an M2 terminated surface. However, the cleavage energy for creating an O-II + SiO-II pair is only slightly larger than that for the M2-M2 pair and thus can also be observed.

A stability diagram for comparing the surface Gibbs free energy for the various possible surface terminations of the forsterite(010) surface has been generated. In the Mg-poor condition, the SiO-II termination has the smallest surface Gibbs free energy. When the environment is Mg-rich, the most stable termination will be M2 or O-II. 
Projections of the phase diagram for ambient and high-temperature conditions have also been generated. The stability ordering of the various terminations does not change as the temperature is increased from room temperature to 2170 K. This calculation is based on the harmonic approximation for vibration, which may, however,not be accurate at such high temperature.  

The calculated results presented here provide a foundation for further theoretical studies of forsterite surfaces, in particular non-stoichiometric surfaces, possible surface reconstructions, and chemical processes that can take place on forsterite(010) surfaces.

\section{\label{sec:level1}Acknowledgements}
This work was supported by 
National Science Foundation of China (Grants \#41503060 and \#41590620), 
Strategic Priority Research Program (B) of Chinese Academy of Sciences (\#XDB18000000 and \#XDB10020301). Ming Geng would like to thank Dr. Javed Husssain is for providing help in VASP calculation and useful discussion in  early stage of this research. Computations were performed on resources provided by the Computer Simulation Lab, IGGCAS and the Computer Center of the University of Iceland.

\bibliography{Manuscript}

\begin{table*}
	\caption{Experimental and Calculated Values for Cell Parameters of Bulk Forsterite (in \AA)}
	\label{tab:LatticeConstant}	
	\begin{tabular}{c|c|c|c|c}		
		\hline \hline
		Lattice Constant	& This work &Experiment \cite{RN298259} & Gaussian Basis \cite{RN298257} &Bruno et al. \cite{RN2646} \\ 
		\hline 
		a   & 4.7654 & 4.756 & 4.804 &4.7892  \\ 
		b	& 10.2396 & 10.207 &10.280  &10.2539  \\ 	
		c	& 5.9984 & 5.980 & 6.032 &  6.0092\\ 
		\hline 	\hline
	\end{tabular}   
\end{table*}

\begin{table*}
	\caption{Calculated Cleavage Energies}	
	\label{tab:CleavageEnergy}
	\begin{tabular}{c|c|c}
		\hline \hline
		Created pairs of surfaces & Cleavage energy (eV/unit cell) & Cleavage energy (J/$m^2$)\\ \hline
		M1 + O &11.765&6.594   \\
		M2 + M2 &3.619&2.029 \\
		M2-II + O2-II &9.841& 5.516 \\
		O-II + SiO-II &5.586&3.130\\
		O2 + SiO &11.904&6.672\\
		\hline \hline
	\end{tabular}
\end{table*}

\begin{table*}
	\caption{Variation in Gibbs free energy for oxygen gas at standard pressure with respect to its value at 0 K. Data are taken from NIST-JANAFtable\cite{RN99999}}	
	 \label{tab:OxygenGibbs}	 
	\begin{tabular}{c|c|c|c|c|c}	
		\hline\hline
		T(K) &$\Delta G^{gas}_{O_2}(T,p^0)(eV)$&$\Delta G^{gas}_O(T,p^0)(eV)$ &T(K)&$\Delta G^{gas}_{O_2}(T,p^0)(eV)$&$\Delta G^{gas}_O(T,p^0)(eV)$ \\
		\hline
		100&       -0.146&-0.073&800 &-1.699&-0.850\\ 
		200&      -0.338&-0.169&900 &-1.945&-0.973\\ 
		298.15& -0.539&-0.270&1000&-2.196&-1.098\\ 
		300&      -0.545&-0.273&1100&-2.450&-1.225\\ 
		400&      -0.762&-0.381&1200&-2.708&-1.354\\ 
		500&      -0.987&-0495&1300&-2.968&-1.493\\ 
		600&      -1.219&-0.610&1400&-3.232&-1.616\\ 
		700&      -1.457&-0.729&1500&-3.498&-1.749\\ \hline	\hline
	\end{tabular}
\end{table*}

\begin{table*}
	\caption{Excesses of O and Mg atoms in slabs with respect to Si atoms and the free energy of formation for various surfaces}
	\label{tab:Excess_atoms}
	\begin{tabular}{c|c|c|c|c|c}
		\hline \hline
		Surface i & $N_{Si}$ &$\Gamma^i_{Si,Mg}$&$\Gamma^i_{Si,O}$ & $\phi ^i (eV/unit cell)$ & $\phi ^i (J/m^2)$ \\
		\hline
		M1-term. &	8&	1&	0&2.645&1.483		\\ 
		M2-term&	8&	0&	0&1.810 &1.014	\\ 
		M2-II-term&	8&	1&	0&2.163 &1.212	\\ 
		O-term&	10&	-1&	0&9.154  &5.131		\\ 
		O-II-term&	8&	1&	1&-3.057&-1.713		\\
		O2-term&	8&	1&	2&-0.839&-0.469	\\
		O2-II-term&10&	-1&	0&9.878&5.537		\\ 
		SiO-term&	10&	-1&	-2& 12.815&	7.186	\\
		SiO-II-term&	10&	-1&	-1& 8.642&	4.844	\\ \hline \hline
	\end{tabular}
\end{table*}

\begin{table*}[]
	\caption{Atom displacement of the topmost layers of nine terminated surfaces ($d_{rel} (\%)$)}
	\label{tab:Displacement}
	\resizebox{\textwidth}{20mm}{
	\begin{tabular}{c|c|c|c|c|c|c|c|c}
		\hline\hline
M1 term.          & M2 term.          & M2-II term.       & O term.          & O-II term.        & O2 term.          & O2-II term.       & SiO term.         & SiO-II term.      \\ \hline
4.45 (M1-O1)      & 4.64 (M2-O2)      & -1.44 (M2-M2)     & -8.73 (O1-SiO)   & -1.75 (O1-M1)     & 0.70 (O2-M2)      & -0.12 (O2 split)  & -3.09 (SiO split) & -3.38 (SiO split) \\
-1.53 (O1-SiO)    & 0.22 (O2-SiO)     & 0.94 (M2-O2)      & 4.15 (SiO-O2)    & 3.84 (M1-O1)      & 1.15 (M2-M2)      & -1.99 (O2-SiO)    & 0.47 (Si-O1)      & 3.01 (Si-O2)      \\
-1.35 (SiO split) & -1.51 (SiO split) & -0.36 (O2 split)  & -5.17 (O2 split) & -1.42 (O1-SiO)    & -0.12 (M2-O2)     & -1.57 (SiO split) & -0.32 (O1-M1)     & -0.78 (O2-M2)     \\
-1.73 (SiO-O2)    & 0.16 (SiO-O1)     & 0.03 (O2-SiO)     & 1.89 (O2-M2)     & -0.23 (SiO split) & -0.13 (O2 split)  & -0.11 (SiO-O1)    & 2.76 (M1-O1)      & 2.33 (M2-M2)      \\
1.20 (O2-M2)      & 0.74 (O1-M1)      & -0.54 (SiO split) & -0.48 (M2-M2)    & -0.66 (SiO-O2)    & -0.25 (O2-SiO)    & 1.55 (O1-M1)      & -1.75 (O1-SiO)    & -1.28 (M2-O)      \\
-2.09 (M2-M2)     & 0.36 (M1-O1)      & -1.16 (SiO-O1)    & -0.61 (M2-O2)    & 1.42 (O2-M2)      & -0.72 (SiO split) & -0.13 (M1 split)  & -0.22 (SiO split) & -0.14 (O2-Si)     \\
0.88 (M2-O2)      & -0.16 (O1-SiO)    & 1.32 (O1-M1)      & -1.60 (O2 split) & -2.27 (M2-M2)     & 0.04 (SiO-O1)     & -1.20 (M1-O1)     & -0.68 (SiO-O2)    & -0.18 (SiO split) \\
-0.05 (O2-SiO)    & -0.99 (SiO split) & -1.67 (M1 split)  & -0.65 (O2-SiO)   & 0.75 (M2-O2)      & 0.27 (O1-M1)      & 0.29 (O1-SiO)     & 0.70 (O2-M2)      & 0.33 (O-O1)       \\
-0.41 (SiO-O1)    & 0.13 (SiO-O2)     & 1.51 (M1-O1)      & -0.71 (SiO-O1)   & -0.51 (O2-SiO)    & 0.34 (M1-O1)      & -0.43 (SiO split) & -0.51 (M2-M2)     & -0.20 (O1-M1)     \\
0.35 (O1-M1)      & 0.62 (O2-M2)      & -0.58 (O1-SiO)    & 0.65 (O1-M1)     & -0.40 (SiO split) & -0.53 (O1-SiO)    & -0.21 (SiO-O2)    & 0.30 (M2-O2)      & 0.07 (M1-O1)       \\ \hline\hline
	\end{tabular}}
\end{table*}

\begin{table*}
	\caption{Bader charges of the bulk crystal and slab with different terminations }
	\label{tab:Bader}
	\resizebox{\textwidth}{20mm}{
\begin{tabular}{cc|cc|cc|cc|cc|cc|cc|cc|cc|cc}
	\hline\hline
	\multicolumn{2}{c|}{Unit Cell} & \multicolumn{2}{c|}{M1 term.} & \multicolumn{2}{c|}{M2 term.} & \multicolumn{2}{c|}{M2-II term.} & \multicolumn{2}{c|}{O term.} & \multicolumn{2}{c|}{O-II term.} & \multicolumn{2}{c|}{O2 term.} & \multicolumn{2}{c|}{O2-II term.} & \multicolumn{2}{c|}{SiO term.} & \multicolumn{2}{c|}{SiO-II term.} \\ \hline
	atom           & charge       & atom           & charge       & atom           & charge       & atom            & charge         & atom          & charge       & atom            & charge        & atom           & charge       & atom            & charge         & atom           & charge        & atom             & charge         \\ \hline
	Mg(M1)         & 1.67         & Mg(M1)         & 1.59         & Mg(M2)         & 1.66         & Mg(M2)          & 0.44           & O(O1)         & -0.72        & O(O1)           & -1.60         & O(O2)          & -0.86        & O(O2)           & -1.06          & Si(SiO)        & 1.58          & O(SiO)           & 1.50           \\
	Mg(M2)         & 1.69         & O(O1)          & -1.60        & O(O2)          & -1.60        & Mg(M2)          & 1.45           & O(SiO)        & -0.89        & Mg(M1)          & 1.64          & O(O2)          & -0.88        & O(O2)           & -1.07          & O(SiO)         & -1.59         & Si(SiO)          & 2.95           \\
	O(O1)          & -1.63        & O(SiO)         & -1.62        & Si(SiO)        & 3.09         & O(O2)           & -1.63          & Si(SiO)       & 3.08         & O(O1)           & -1.59         & Mg(M2)         & 1.67         & O(SiO)          & -1.49          & O(O1)          & -1.59         & O(O2)            & -1.56          \\
	O(O2)          & -1.61        & Si(SiO)        & 1.62         & O(SiO)         & -1.58        & O(O2)           & -1.70          & O(O2)         & -1.56        & O(SiO)          & -1.62         & Mg(M2)         & 1.68         & Si(SiO)         & 3.07           & Mg(M1)         & 1.66          & Mg(M2)           & 1.68           \\
	O(SiO)         & -1.62        & O(O2)          & -1.61        & O(O1)          & -1.62        & O(SiO)          & -1.66          & O(O2)         & -1.58        & Si(SiO)         & 3.08          & O(O2)          & -1.58        & O(O1)           & -1.35          & O(O1)          & -1.63         & O(O2)            & -1.61          \\
	Si(SiO)        & 3.1          & Mg(M2)         & 1.68         & Mg(M1)         & 1.66         & Si(SiO)         & 3.09           & Mg(M2)        & 1.68         & O(O2)           & -1.61         & Si(SiO)        & 3.08         & Mg(M1)          & 1.67           & O(SiO)         & -1.62         & Si(SiO)          & 3.08           \\
	&              & Mg(M2)         & 1.68         & O(O1)          & -1.61        & O(O1)           & -1.64          & O(O2)         & -1.59        & Mg(M2)          & 1.69          & O(SiO)         & -1.60        & O(O1)           & -1.55          & Si(SiO)        & 3.08          & O(SiO)           & -1.61          \\
	&              & O(O2)          & -1.61        & O(SiO)         & -1.61        & Mg(M1)          & 1.66           & O(O2)         & -1.62        & Mg(M2)          & 1.68          & O(O1)          & -1.61        & O(SiO)          & -1.59          & O(O2)          & -1.61         & O(O1)            & -1.62          \\
	&              & Si(SiO)        & 3.08         & Si(SiO)        & 3.08         & O(O1)           & -1.62          & Si(SiO)       & 3.10         & O(O2)           & -1.60         & Mg(M1)         & 1.66         & Si(SiO)         & 3.08           & Mg(M2)         & 1.68          & Mg(M1)           & 1.66           \\
	&              & O(SiO)         & -1.60        & O(O2)          & -1.60        & O(SiO)          & -1.60          & O(SiO)        & -1.62        & O(SiO)          & -1.59         & O(O1)          & -1.61        & O(O2)           & -1.54          & O(O2)          & -1.6          & O(O1)            & -1.62         \\\hline\hline
\end{tabular}}
\end{table*}

\begin{figure*}[h]
	\centering
	\includegraphics[width=0.9\linewidth]{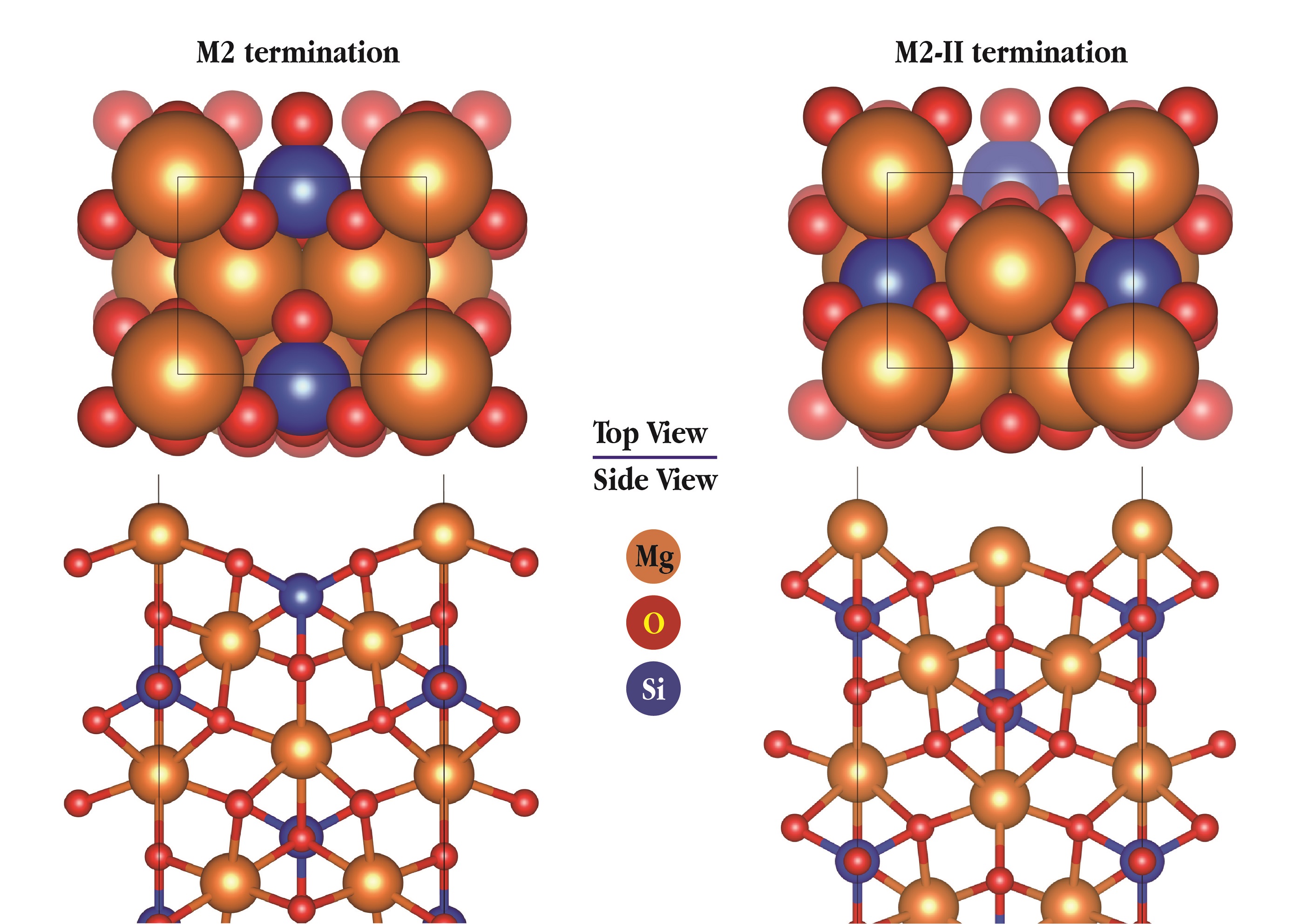}
	\caption{M2 termination surface structure}
	\label{fig:M2}
\end{figure*}

\begin{figure*}[h]
	\centering
	\includegraphics[width=0.9\linewidth]{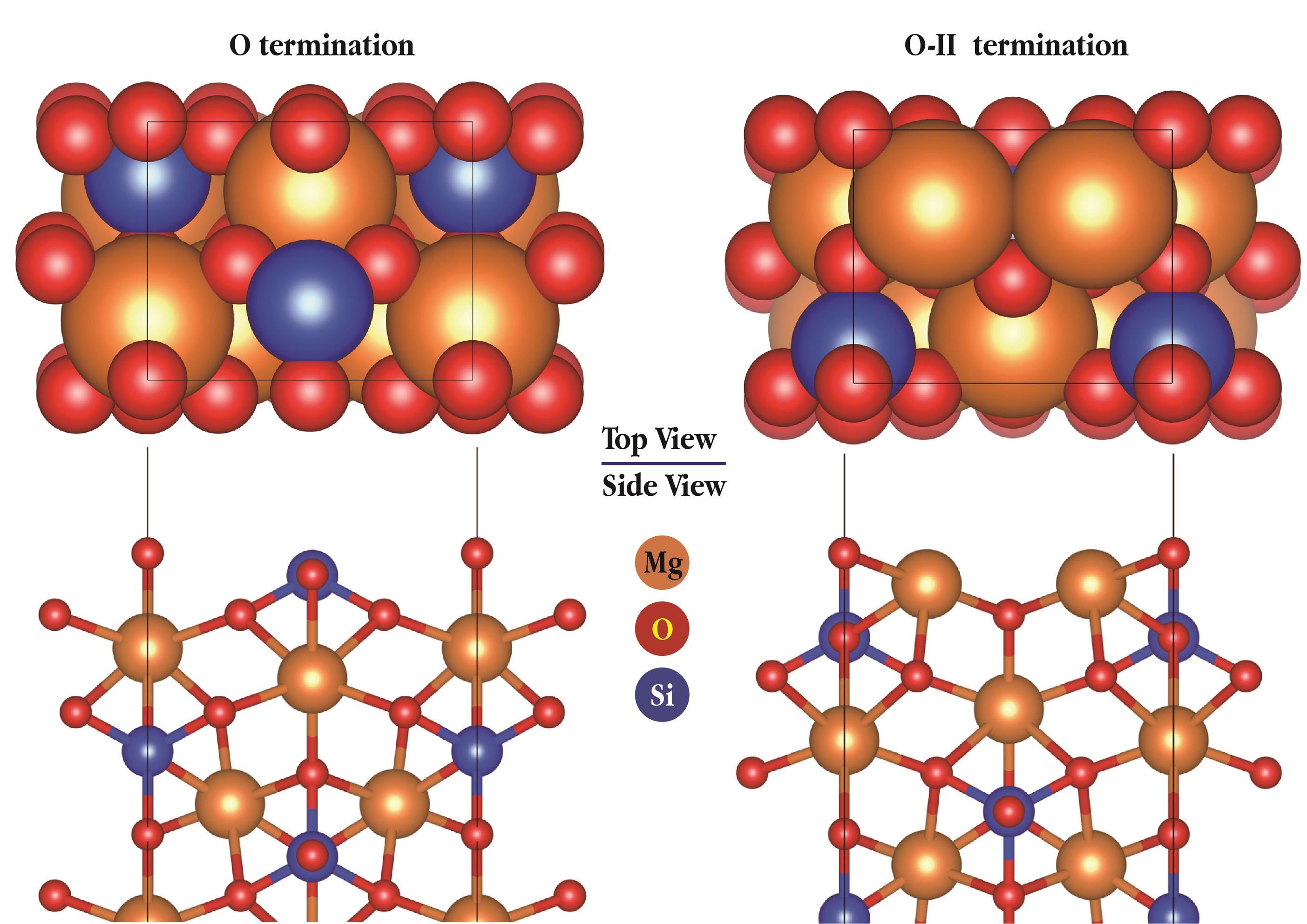}
	\caption{O termination surface structure}
	\label{fig:O}
\end{figure*}

\begin{figure*}[h]
	\centering
	\includegraphics[width=0.9\linewidth]{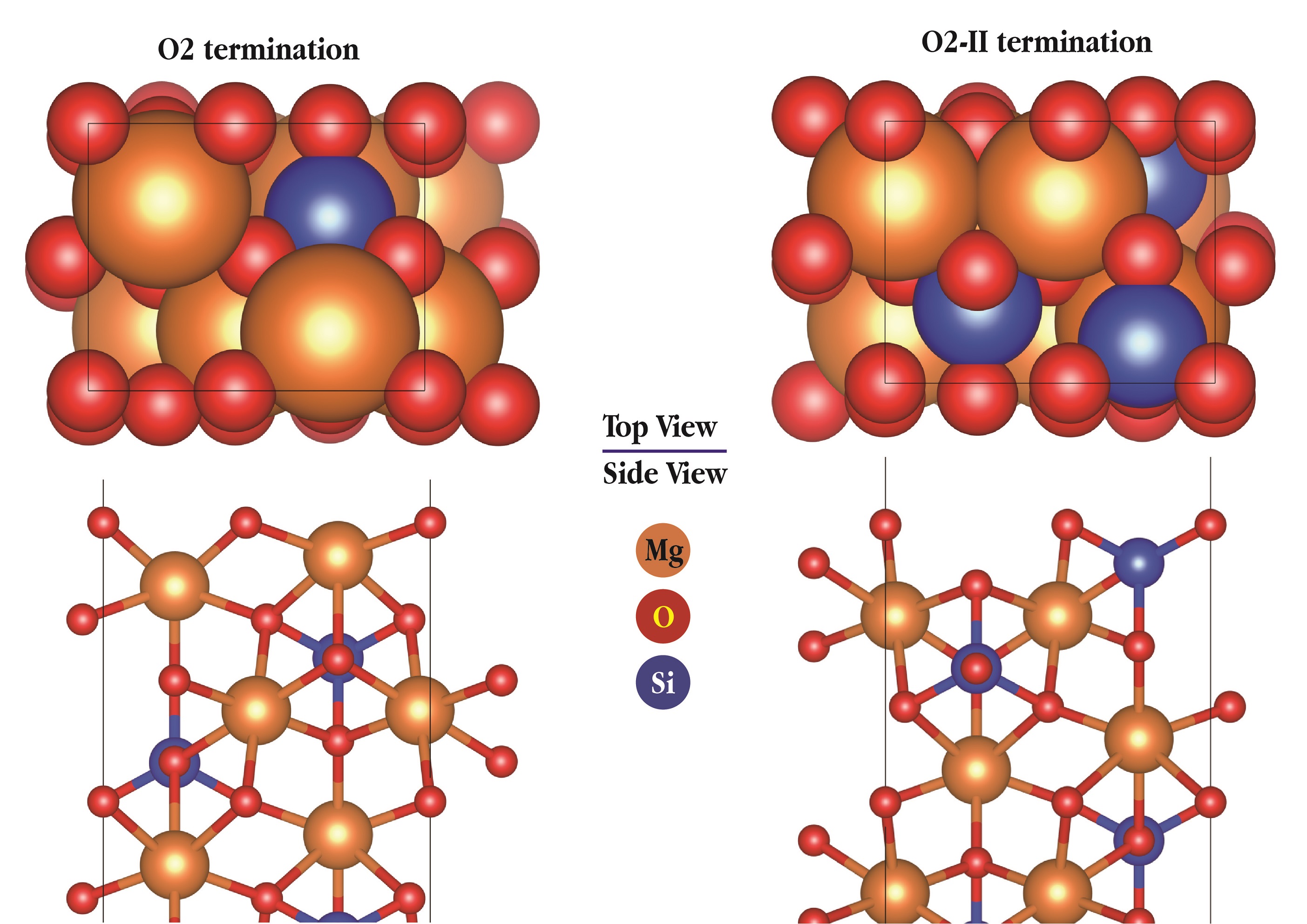}
	\caption{O2 termination at the surface structure}
	\label{fig:O2}
\end{figure*}

\begin{figure*}[h]
	\centering
	\includegraphics[width=0.9\linewidth]{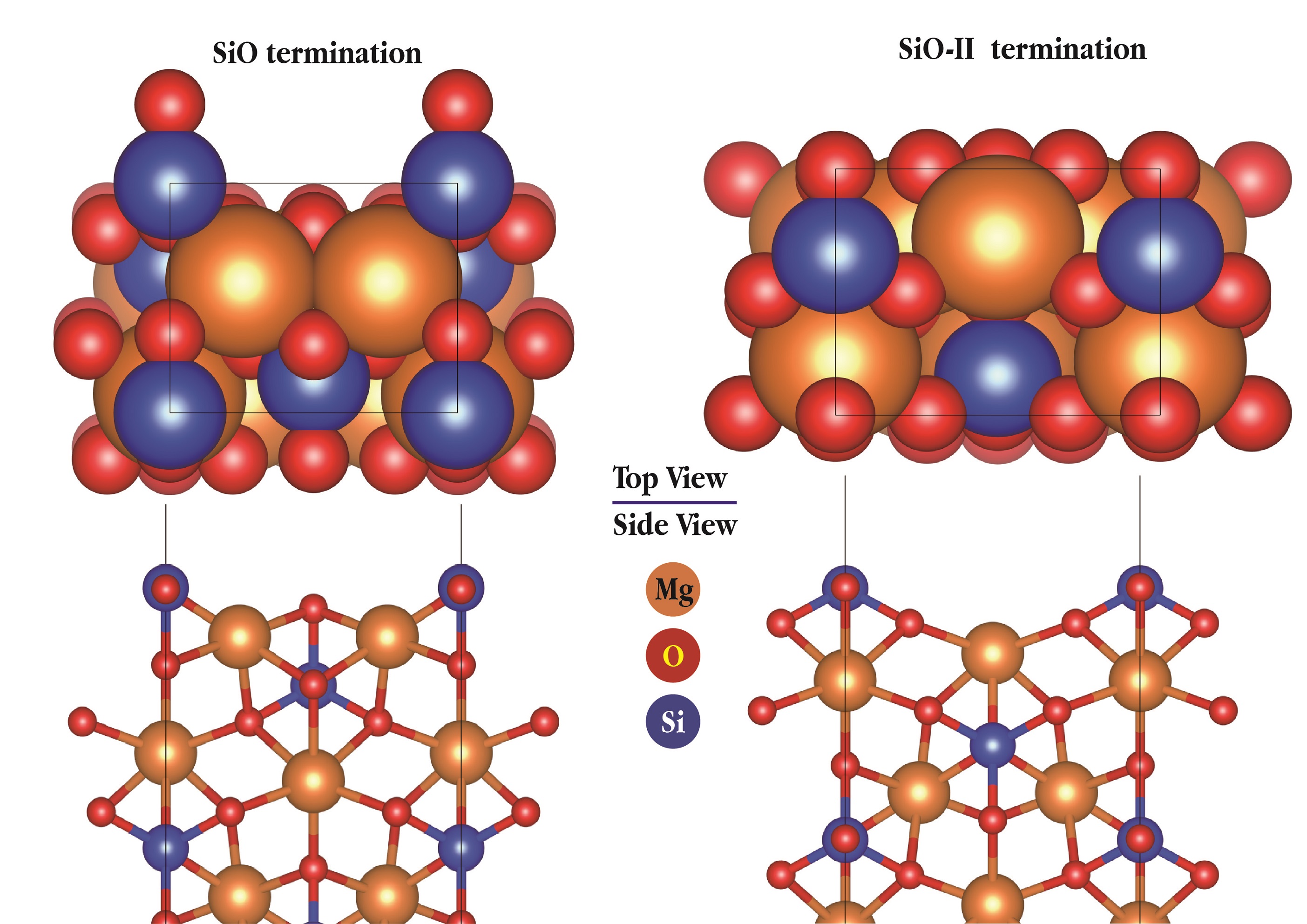}
	\caption{SiO termination at the surface structure}
	\label{fig:SiO}
\end{figure*}

\end{document}